# Layer dependence of stacking order in nonencapsulated few-layer CrI$_3$

Kai Guo[1,4], Bowen Deng[3], Zhen Liu[1,4], Chaofeng Gao[2], Zhongtai Shi[1,4], Lei Bi[1,4], Li Zhang[1,4], Haipeng Lu[1,4], Peiheng Zhou[1,4], Linbo Zhang[1,4,*], Yingchun Cheng[2] and Bo Peng[1,4,*]

[1] National Engineering Research Center of Electromagnetic Radiation Control Materials, School of Electronic Science and Engineering, University of Electronic Science and Technology of China, Chengdu 611731, China

[2] Key Laboratory of Flexible Electronics & Institute of Advanced Materials, Jiangsu National Synergetic Innovation Center for Advanced Materials, Nanjing Tech University, Nanjing 211816, China

[3] Department of Physics, Broida Hall, University of California Santa Barbara, CA 93106-9530

[4] Key Laboratory of Multi-spectral Absorbing Materials and Structures of Ministry of Education, University of Electronic Science and Technology of China, Chengdu 611731, China

* To whom correspondence should be addressed. Email address: bo_peng@uestc.edu.cn; zhanglinbo@uestc.edu.cn

**Abstract:** Long-range magnetic orders in atomically thin ferromagnetic $CrI_3$ give rise to new fascinating physics and application perspectives. The physical properties of two-dimensional (2D) ferromagnetism $CrI_3$ are significantly influenced by interlayer spacing and stacking order, which are sensitive to the hydrostatic pressure and external environments. However, there remains debate on the stacking order at low temperature. Here, we study the interlayer coupling and stacking order of non-encapsulated 2-5 layer and bulk $CrI_3$ at 10 K by Raman spectroscopy; demonstrate a rhombohedral stacking in both antiferromagnetic and ferromagnetic $CrI_3$. The opposite helicity dependence of $A_g$ and $E_g$ modes arising from phonon symmetry further validate the rhombohedral stacking. An anomalous temperature-dependent behavior is observed due to spin-phonon coupling below 60 K. Our work provides insights into the interlayer coupling and stacking orders of 2D ferromagnetic materials.

Since the discovery of two ferromagnetic (FM) atomically thin $CrI_3$ and $Cr_2Ge_2Te_6$ in 2017, intrinsic ferromagnetism in two-dimensional (2D) van der Waals (vdWs) materials, maintaining long-range magnetic orders at the atomic monolayer limit, has received growing attention[1-7]. 2D vdW ferromagnetic materials as spin-filters have been integrated into van der Walls heterostructures, demonstrating giant tunneling magnetoresistance towards the next-generation information transfer and data storage technologies[8-10]. Among 2D ferromagnetic materials, $CrI_3$ is of particular interest. Bulk $CrI_3$ is FM below the Curie temperature ($T_c$), but few-layer $CrI_3$ display striking layer-dependent magnetism. Each individual layer is FM, however adjacent layers are antiferromagnetically coupled together and becomes a layered antiferromagnet (AFM) when thinned down to a few atomic layers[1]. The interlayer magnetic state can be switched between FM and AFM by electric gating or electrostatic doping[11-14] and pressure[15, 16].

In a vdWs material and heterostructures, a tiny change of lattice constant and interlayer coupling between adjacent layers can drastically influence their physical properties. Bulk $CrI_3$ adopts the monoclinic stacking at room temperature while transfer to rhombohedral stacking at ~210 K, bulk and encapsulated few-layer $CrI_3$ with rhombohedral stacking has been reported to be ferromagnetic below ~61 K[17, 18]. However, recent experiments demonstrate that the BN-encapsulated bi- and few-layer $CrI_3$ and $CrCl_3$ belong to monoclinic structure (point group $C_{2h}$ (2/m)) rather than rhombohedral structure at low temperature[19, 20], and some theoretical proposals demonstrate that the antiferromagnetic coupling is associated with monoclinic layer stacking[21]. Thus, there still is a debate on the stacking order of CrI$_3$ at low temperature.

A complete understanding of the low-temperature lattice structure and stacking order is crucial for 2D vdW ferromagnetic materials. Raman spectroscopy is a powerful tool to study the crystal structures, lattice vibration and shift and interlayer coupling[22-28]. In this work, we study the layer, polarization and helicity dependencies of Raman features of non-encapsulated 2-5 layer and bulk CrI$_3$ at 10 K, demonstrating that few-layer and bulk CrI$_3$ are rhombohedral phase at low temperature, including antiferromagnetic bilayer CrI$_3$. The peak intensities of $E_g^3$ and $E_g^4$ mode are independent of the polarization angle, while the intensity of $A_g^2$ and $A_g^3$ show a two-fold polarization dependence. The polarization-resolved Raman spectra demonstrate that the $A_g^3$ and $E_g^4$ features show opposite helicities in which the $A_g^3$ mode maintains the helicity of incident light, however, the $E_g^4$ mode reverses it. Under cooling, the $A_g^3$ and $E_g^4$ modes shift to higher frequencies; remarkably, a spin-phonon coupling takes place and causes an anomalous behavior for the Raman feature below ~60 K.

**RESULTS AND DISCUSSION**

**Layer dependence of magnetic order.** In 2D ferromagnetic CrI$_3$, the Cr$^{3+}$ ions in each layer are coordinated by six nonmagnetic I$^-$ ions to form an octahedral geometry, which further share edges to build a honeycomb network (Fig. 1a). Bulk CrI$_3$ crystals undergo a phase transition to a rhombohedral structure (space group $R\overline{3}$, Fig. 1b) at ~210-220 K from a monoclinic structure (space group C2/m, Fig. 1c) at room temperature[17]. Figure 1d shows the non-encapsulated $CrI_3$ with 2 to 6 layers on $SiO_2$ (300 nm)/Si substrates, which are in-suit loaded into cold head with an optical window in a glovebox. The different numbers of layers can be identified through the optical contrast. Our experimental optical contrast results shown in Fig. 1e (red circles) are consistent with the reported experimental and calculated results from Xu *et al*. (solid blue line in Fig. 1e)[1]. Bulk $CrI_3$ crystals show distinct ferromagnetism below ~65 K (Fig. 1f).

Figure 2 show the reflective magnetic circular dichroism (RMCD) signal in atomically-thin $CrI_3$ at 10 K. RMCD signals from a bilayer $CrI_3$ flake approach zero for applied magnetic fields ± 0.65 T, indicating antiferromagnetic behavior (Fig. 2a). In bilayer $CrI_3$, the two layers have opposite magnetic orientation and nearly compensate for the intrinsic magnetocrystalline anisotropy each other, therefore, the net magnetization disappear and 2L $CrI_3$ is antiferromagnetic[1, 2]. However, striking hysteresis is observed in 3L and 4L $CrI_3$ flakes (Fig. 2b and 2c). In 3L $CrI_3$, the magnetizations of the three layers are oriented to the same direction, while the magnetization of one layer is opposite with the other three in 4L $CrI_3$. Thus, the net magnetization of 3L $CrI_3$ is stronger than that of 4L $CrI_3$.

**Layer dependence of stacking order.** The stacking order and crystal structures of non-encapsulated few-layer $CrI_3$ can be identified by polarized Raman spectroscopy. The polarization selection rules were performed on a 2-5 layer (2L, 3L, 4L and 5L) and bulk $CrI_3$ at 10 K using a confocal Raman microscope system with a backscattering geometry (Fig. 3a-d). The incident and scattered light are along the $-z$ and $+z$ direction, respectively. The polarization configuration of the incident and scattered light are parallel (XX) or perpendicular (XY) to each other. The feature at ~103 and 128 cm$^{-1}$ is only present in the parallel configuration (XX), while the Raman peak at ~107 and ~238 cm$^{-1}$ appear for both parallel and perpendicular configurations (XX and XY). The Raman scattering intensities $I$ are proportional to $|e_s \cdot R \cdot e_i|^2$, where the $e_i$ and $e_s$ are the polarization unit vectors of the incident and scattered light, respectively[26, 29, 30]. The $R$ is the Raman tensor of the Raman-active vibrational modes as predicted by group theory.

We consider both the rhombohedral (point group, $C_{3i}$) and monoclinic phase (point group, $C_{2h}$). The scattered light polarization has an angle of $\theta$ with that of the incident light, which is parallel to the x-axis. Thus, $\hat{e}_i = (1, 0, 0)$ and $\hat{e}_s = (\cos\theta, \sin\theta, 0)$, and $\theta = 0°$ and $90°$ under the XX and XY polarization configuration.

In rhombohedral structure, Raman active modes are the $A_g$ and $E_g$ modes and corresponding Raman tensors are given by[31]

$$A_g(R): \begin{pmatrix} a & 0 & 0 \\ 0 & a & 0 \\ 0 & 0 & b \end{pmatrix}, \quad {}^1E_g(R): \begin{pmatrix} c & d & e \\ d & -c & f \\ e & f & 0 \end{pmatrix}, \quad {}^2E_g(R): \begin{pmatrix} d & -c & -f \\ -c & -d & e \\ -f & e & 0 \end{pmatrix}$$

Taking $e_i$, $e_s$ and $R$ to $I \propto |e_s \cdot R \cdot e_i|^2$, the Raman intensities of $A_g$ and $E_g$ are

$$I_{A_g}(R) \propto a^2 \cos^2\theta$$

$${}^1I_{E_g}(R) \propto c^2 \cos^2\theta + cd\sin(2\theta) + d^2 \sin^2\theta,$$

$${}^2I_{E_g}(R) \propto d^2 \cos^2\theta - cd\sin(2\theta) + c^2 \sin^2\theta$$

The $E_g$ modes are degenerated, thus all $E_g$ modes contribute to a peak at the same frequency, $I_{E_g}(R) = {}^1I_{E_g}(R) + {}^2I_{E_g}(R) = c^2 + d^2$, giving a constant. Thus, the $E_g$ modes are independent of the polarization, which can be detected in any polarization configuration. In contrast, the $A_g$ modes show striking two-fold polarization dependence, which can be observed only in parallel polarization configuration.

In monoclinic structure, the symmetry is lowered and $E_g$ modes split to an $A_g$ and a $B_g$ mode. Thus, only the $A_g$ and $B_g$ modes are active and corresponding Raman tensors are given by

$$A_g(M): \begin{pmatrix} a & 0 & d \\ 0 & b & 0 \\ d & 0 & c \end{pmatrix}, \quad B_g(M): \begin{pmatrix} 0 & e & 0 \\ e & 0 & f \\ 0 & f & 0 \end{pmatrix}$$

The Raman intensities of $A_g$ and $B_g$ are

$$I_{A_g}(M) \propto a^2 \cos^2\theta, \quad I_{B_g}(M) \propto e^2 \sin^2\theta$$

Thus, the Raman intensities of $A_g$ and $B_g$ modes have opposite two-fold dependence on polarization angle in the monoclinic phase.

Therefore, group theory predicts that the Raman scattering intensity is zero for $A_g$ modes under the perpendicular (XY) configuration in both the rhombohedral and monoclinic stacking[30]. However, only for rhombohedral phase, there remain the polarization-independent $E_g$ modes that have nonzero Raman intensities under the parallel (XX) and perpendicular (XY) configuration. The polarization angle dependences of the features at ~103, ~107 ~128 and ~234 cm$^{-1}$ of non-encapsulated 2-5 layer and bulk CrI$_3$ are shown in Fig. 3e. The Raman intensities of the peaks near 103 and 128 cm$^{-1}$ have a two-fold pattern versus polarization angle, in contrast, the modes at ~107 and ~234 cm$^{-1}$ are independent with polarization angle. Therefore, the non-encapsulated 2-5 layer and bulk CrI$_3$ are rhombohedral phase rather than monoclinic phase; and the peaks at ~103, ~107 ~128 and ~234 cm$^{-1}$ are assigned to the $A_g^2$, $E_g^3$, $A_g^3$ and $E_g^4$ modes, respectively[18].

To further validate the rhombohedral stacking of CrI$_3$ at low temperature, the helicity selection rules are studied at 10 K (Fig. 4). Figure 4a shows a schematic of the experimental optical setup. The linearly polarized light passes through a 1/4λ waveplate and transfers to circularly polarized incident $e_i = E_0 e^{i(kz-wt)}(\hat{x} \pm i\hat{y})$. The incident and scattered light are along opposite direction, thus, the scattered light $e_s$ are $E_0 e^{i(kz-wt)}(\hat{x} \mp i\hat{y})$ and $E_0 e^{i(kz-wt)}(\hat{x} \pm i\hat{y})$ under the same ($\sigma+\sigma+$) and opposite ($\sigma+\sigma-$) circular polarization configuration, respectively.

The Raman intensities of the $A_g$ mode in rhombohedral phase are $I_{A_g}^{\sigma+\sigma+}(R) \propto a^2$ and $I_{A_g}^{\sigma+\sigma-}(R) \propto 0$, on the contrary, for $E_g$ modes, $I_{E_g}^{\sigma+\sigma+}(R) \propto 0$ and $I_{E_g}^{\sigma+\sigma-}(R) \propto c^2 + d^2$, indicating that the $A_g$ modes maintain the helicity of incident light, however, the $E_g$ modes reverse it. In contrast, for monoclinic phase with lower symmetry, both the $A_g$ and $B_g$ modes reverse the helicity of incident light, in which the Raman intensities of $A_g$ and $B_g$ modes are given by $I_{A_g}^{\sigma+\sigma+}(M) \propto 0$, $I_{A_g}^{\sigma+\sigma-}(M) \propto a^2$, $I_{B_g}^{\sigma+\sigma+}(M) \propto 0$ and $I_{B_g}^{\sigma+\sigma-}(M) \propto e^2$.

The polarization-resolved Raman spectra were obtained under excitation by a left-handed circularly polarized light at 2.41 eV (σ$^+$). Remarkably, the Raman features of 2L, 3L and 4L CrI$_3$ at ~128 cm$^{-1}$ have the same helicity as the incident photon, whereas the peaks at ~234 cm$^{-1}$ switch the helicity (Fig. 4b and 4c). The striking opposite polarization dependence of helicities of the two Raman modes is shown in Fig. 4d. The experimental data are in agreement with the predicted rhombohedral phase results. The helicity-dependent experimental results further validate that the peaks at ~128 and ~234 cm$^{-1}$ are assigned to the $A_g^3$ and $E_g^4$ modes, respectively, consistent with above polarization-resolved Raman results. With increasing magnetic field, the helicity selection behaviors show no detectable changes (Fig. 4c), indicating that this is independent on the external magnetic field. The helicity selection behaviors are only attributed to the symmetric Raman tensors. Thus, 2-5 layer and bulk 2D ferromagnetic CrI$_3$ have rhombohedral stacking order.

**Temperature dependence of the Raman features.** To better understand the lattice dynamics, the temperature dependence of the Raman features was determined in 2L, 3L and 6L CrI$_3$ flakes. Since the transition from the monoclinic phase to the rhombohedral phase occurs at around 210-220 K in CrI$_3$[17], we focus on investigating the Raman features below 200 K. Figure 5a-c show the normalized Raman intensity maps of the 2L, 3L and 6L CrI$_3$ with decreasing temperatures from 200 to 10 K. The $A_g^3$ and $E_g^4$ modes simultaneously blueshift to higher frequencies with decreasing temperatures, which are attributed to the phonon anharmonic decay and lattice contraction. The shifts of the $A_g^3$ modes are ~1.4, ~0.9 and ~0.8 cm$^{-1}$ for the 2L, 3L and 6L CrI$_3$, and the $E_g^4$ modes show shifts of ~2.9, ~2.9 and ~2.4 cm$^{-1}$, respectively.

The linewidths for the $A_g^3$ and $E_g^4$ modes are extracted to investigate the spin-phonon coupling behavior. Figure 5d-i show the linewidths of the $A_g^3$ and $E_g^4$ modes of 2L, 3L and 6L CrI$_3$ with decreasing temperature from 200 to 10 K. The temperature dependence of the linewidths is predicted by phonon-phonon coupling[32], which is given by $\Gamma_L(T) = \Gamma_{L,0}(1 + 2\lambda_{ph\text{-}ph}/(\exp(hC\omega_0/2k_BT)-1))$, where the $\Gamma_{L,0}$ and $\omega_0$ are the zero temperature limits of the linewidth and phonon energy, obtained by extrapolating the experimental results to 0 K. The $\lambda_{ph\text{-}ph}$ represents phonon-phonon coupling constant. Above 60 K (~$T_c$), the linewidths of the $A_g^3$ and $E_g^4$ modes monotonically decrease as the temperature decreases (solid lines in Fig. 5d-i). However, the linewidths deviate from the expected tendency below 60 K. In particular, the $A_g^3$ ($E_g^4$) modes of the 2L (6L) CrI$_3$ exhibit distinct deviation behavior (Fig. 5d-i), which indicates that a new scattering mechanism contributes to the anomalous phonon behavior. The $T_c$ of few-layer CrI$_3$ is ~65 K (Fig. 1f)[3]. The emerging long-range magnetic order is anticipated to result in the abrupt change in the linewidth originating from spin-phonon coupling. We briefly discuss the mechanism of spin-phonon coupling. The total Hamiltonian of lattice in crystals is expressed as $H = H_{ion} + H_{electron} + H_{e-ph}$. In the presence of magnetic orders, an additional Hamiltonian term $H_{spin} = -\sum_{i,j} J_{ij} \langle S_i S_j \rangle$ is introduced, where $J_{ij}$ is the spin-phonon coupling constant and $\langle S_i S_j \rangle$ is spin correlation function[33]. In 2D layered CrI$_3$ materials, the electron-phonon coupling is not observed, thus the Hamiltonian $H_{e-ph}$ is neglected and only phonon-phonon coupling is considered. Thus, the total potential energy of a Raman active mode is consisted of "lattice" and "spin" contributions; corresponding Hamiltonian of Raman modes is expressed as $H = H_{ion} + H_{electron} + H_{spin} = H_{lattice} + H_{spin}$. Therefore, the spin-spin exchange interaction takes place in the ferromagnetic phase at low temperature, resulting in the deviation of linewidth.

**CONCLUSION**

In summary, we demonstrated a comprehensive understanding of stacking order of 2D ferromagnetic CrI$_3$, including layer, polarization and temperature dependence of $A_g$ and $E_g$ modes; demonstrate that the non-encapsulated 2-5 layer and bulk CrI$_3$ are rhombohedral stacking order at 10 K, rather than monoclinic structure. The spin-phonon coupling occurs below ~60 K, resulting in the deviation of linewidth. Zero-momentum spin wave features close to the $A_g$ mode in frequency have been observed.[34] This work highlights the potential to manipulate spin waves through spin-phonon and magnetoelectric coupling in new ferromagnetic 2D materials to produce novel spintronic devices[35].

**Experimental Methods**

**Sample preparation:** Few-layer CrI$_3$ were mechanically exfoliated from a bulk crystal onto PDMS films and were then directly transferred onto SiO$_2$/Si substrates, which were then loaded into the cold head for optical measurements in a glove box.

**Optical measurements:** The Raman signals were recorded using a Witec Alpha 300R Plus confocal Raman microscope with a closed cycle optical cryostat (10 K) and a 7T magnetic field. A power-stabilized 633 nm HeNe laser is modulated by photoelastic modulator (PEM) and coupled to the Witec Raman system. The modulated beam is directed through a non-polarizing beamsplitter cubes to the sample, housed in a closed-cycle cryostat at 10 K. An out-of-plane magnetic field is applied in Faraday geometry. The reflected beam passes through the same non-polarizing beamsplitter cubes, non-PM multimode fiber onto a photodetector, where lock-in detection measures the reflected intensity at $f_{PEM}$ (50 KHz). A long working distance 50× objective (NA = 0.45) was used for the Raman and RMCD measurements. The Raman signals were coupled into the spectrometer with an 1800 and 600 g/mm grating. 1/4λ waveplate and polarization analyzer were used for the polarization-resolved Raman measurements. The power of a 514 nm laser was measured to be approximately 2 mW, and the typical integration time was 30 s.

**References:**

1. Huang, B.; Clark, G.; Navarro-Moratalla, E.; Klein, D. R.; Cheng, R.; Seyler, K. L.; Zhong, D.; Schmidgall, E.; McGuire, M. A.; Cobden, D. H.; Yao, W.; Xiao, D.; Jarillo-Herrero, P.; Xu, X. Layer-dependent ferromagnetism in a van der waals crystal down to the monolayer limit. *Nature* **2017**, *546*, 270-273.

2. Gong, C.; Li, L.; Li, Z.; Ji, H.; Stern, A.; Xia, Y.; Cao, T.; Bao, W.; Wang, C.; Wang, Y.; Qiu, Z. Q.; Cava, R. J.; Louie, S. G.; Xia, J.; Zhang, X. Discovery of intrinsic ferromagnetism in two-dimensional van der waals crystals. *Nature* **2017**, *546*, 265-269.

3. Zhong, D.; Seyler, K. L.; Linpeng, X.; Cheng, R.; Sivadas, N.; Huang, B.; Schmidgall, E.; Taniguchi, T.; Watanabe, K.; McGuire, M. A.; Yao, W.; Xiao, D.; Fu, K.-M. C.; Xu, X. Van der waals engineering of ferromagnetic semiconductor heterostructures for spin and valleytronics. *Sci. Adv.* **2017**, *3*, e1603113.

4. Deng, Y.; Yu, Y.; Song, Y.; Zhang, J.; Wang, N. Z.; Sun, Z.; Yi, Y.; Wu, Y. Z.; Wu, S.; Zhu, J.; Wang, J.; Chen, X. H.; Zhang, Y. Gate-tunable room-temperature ferromagnetism in two-dimensional fe$_3$gete$_2$. *Nature* **2018**, *563*, 94-99.

5. Mounet, N.; Gibertini, M.; Schwaller, P.; Campi, D.; Merkys, A.; Marrazzo, A.; Sohier, T.; Castelli, I. E.; Cepellotti, A.; Pizzi, G.; Marzari, N. Two-dimensional materials from high-throughput computational exfoliation of experimentally known compounds. *Nat. Nanotechnol.* **2018**, *13*, 246-252.

6. Lee, J.-U.; Lee, S.; Ryoo, J. H.; Kang, S.; Kim, T. Y.; Kim, P.; Park, C.-H.; Park, J.-G.; Cheong, H. Ising-type magnetic ordering in atomically thin FePS$_3$. *Nano Lett.*

**2016**, *16*, 7433-7438.

7. Bonilla, M.; Kolekar, S.; Ma, Y.; Diaz, H. C.; Kalappattil, V.; Das, R.; Eggers, T.; Gutierrez, H. R.; Phan, M.-H.; Batzill, M. Strong room-temperature ferromagnetism in vse$_2$ monolayers on van der waals substrates. *Nat. Nano.* **2018**, *13*, 289-293.

8. Song, T.; Cai, X.; Tu, M. W.-Y.; Zhang, X.; Huang, B.; Wilson, N. P.; Seyler, K. L.; Zhu, L.; Taniguchi, T.; Watanabe, K.; McGuire, M. A.; Cobden, D. H.; Xiao, D.; Yao, W.; Xu, X. Giant tunneling magnetoresistance in spin-filter van der waals heterostructures. *Science* **2018**, *360*, 1214-1218.

9. Klein, D. R.; MacNeill, D.; Lado, J. L.; Soriano, D.; Navarro-Moratalla, E.; Watanabe, K.; Taniguchi, T.; Manni, S.; Canfield, P.; Fernández-Rossier, J.; Jarillo-Herrero, P. Probing magnetism in 2d van der waals crystalline insulators via electron tunneling. *Science* **2018**, *360*, 1218-1222.

10. Wang, Z.; Gutiérrez-Lezama, I.; Ubrig, N.; Kroner, M.; Gibertini, M.; Taniguchi, T.; Watanabe, K.; Imamoğlu, A.; Giannini, E.; Morpurgo, A. F. Very large tunneling magnetoresistance in layered magnetic semiconductor cri$_3$. *Nat. Commun.* **2018**, *9*, 2516.

11. Huang, B.; Clark, G.; Klein, D. R.; MacNeill, D.; Navarro-Moratalla, E.; Seyler, K. L.; Wilson, N.; McGuire, M. A.; Cobden, D. H.; Xiao, D.; Yao, W.; Jarillo-Herrero, P.; Xu, X. Electrical control of 2D magnetism in bilayer CrI$_3$. *Nat. Nanotechnol.* **2018**, *13*, 544-548.

12. Jiang, S.; Shan, J.; Mak, K. F. Electric-field switching of two-dimensional van der waals magnets. *Nat. Mater.* **2018**, *17*, 406-410

13. Jiang, S.; Li, L.; Wang, Z.; Mak, K. F.; Shan, J. Controlling magnetism in 2D CrI$_3$ by electrostatic doping. *Nat. Nanotechnol.* **2018**, *13*, 549-553.

14. Wang, Z.; Zhang, T.; Ding, M.; Dong, B.; Li, Y.; Chen, M.; Li, X.; Huang, J.; Wang, H.; Zhao, X.; Li, Y.; Li, D.; Jia, C.; Sun, L.; Guo, H.; Ye, Y.; Sun, D.; Chen, Y.; Yang, T.; Zhang, J.*, et al.* Electric-field control of magnetism in a few-layered van der waals ferromagnetic semiconductor. *Nat. Nanotechnol.* **2018**, *13*, 554-559.

15. Li, T.; Jiang, S.; Sivadas, N.; Wang, Z.; Xu, Y.; Weber, D.; Goldberger, E. J.; Watanabe, K.; Taniguchi, T.; Fennie, J. C.; Mak, K. F.; Shan, J. Pressure-controlled interlayer magnetism in atomically thin CrI$_3$. *arXiv:1905.10905* **2019**.

16. Song, T.; Fei, Z.; Yankowitz, M.; Lin, Z.; Jiang, Q.; Hwangbo, K.; Zhang, Q.; Sun, B.; Taniguchi, T.; Watanabe, K.; McGuire, A. M.; Graf, D.; Cao, T.; Chu, J.-H.; Cobden, D. H.; Dean, C. R.; Xiao, D.; Xu, X. Switching 2d magnetic states via pressure tuning of layer stacking. *arXiv:1905.10860* **2019**.

17. McGuire, M. A.; Dixit, H.; Cooper, V. R.; Sales, B. C. Coupling of crystal structure and magnetism in the layered, ferromagnetic insulator CrI$_3$. *Chem. Mater.* **2015**, *27*, 612-620.

18. Djurdjić-Mijin, S.; Šolajić, A.; Pešić, J.; Šćepanović, M.; Liu, Y.; Baum, A.; Petrovic, C.; Lazarević, N.; Popović, Z. V. Lattice dynamics and phase transition in CrI$_3$ single crystals. *Phys. Rev. B* **2018**, *98*, 104307.

19. Sun, Z.; Yi, Y.; Song, T.; Clark, G.; Huang, B.; Shan, Y.; Wu, S.; Huang, D.; Gao, C.; Chen, Z.; McGuire, M.; Cao, T.; Xiao, D.; Liu, W.-T.; Yao, W.; Xu, X.; Wu, S. Giant nonreciprocal second-harmonic generation from antiferromagnetic bilayer CrI$_3$. *Nature* **2019**, *572*, 497-501.

20. Klein, D. R.; MacNeill, D.; Song, Q.; Larson, D. T.; Fang, S.; Xu, M.; Ribeiro, R. A.; Canfield, P. C.; Kaxiras, E.; Comin, R.; Jarillo-Herrero, P. Enhancement of interlayer exchange in an ultrathin two-dimensional magnet. *Nat. Phys.* **2019,** doi.org/10.1038/s41567-019-0651-0.

21. Sivadas, N.; Okamoto, S.; Xu, X.; Fennie, C. J.; Xiao, D. Stacking-dependent magnetism in bilayer CrI$_3$. *Nano. Lett.* **2018**, *18*, 7658-7664.

22. Webster, L.; Liang, L.; Yan, J.-A. Distinct spin–lattice and spin–phonon interactions in monolayer magnetic CrI$_3$. *Phys. Chem. Chem. Phys.* **2018**, *20*, 23546-23555.

23. Sun, L.; Zheng, J., Optical visualization of MoS$_2$ grain boundaries by gold deposition. *Sci. China. Mater.* **2018,** *61*, 1154-1158.

24. Wu, S.; Shi, X.; Liu, Y.; Wang, L.; Zhang, J.; Zhao, W.; Wei, P.; Huang, W.; Huang, X.; Li, H., The influence of two-dimensional organic adlayer thickness on the ultralow frequency raman spectra of transition metal dichalcogenide nanosheets. *Sci. China. Mater.* **2019,** *62*, 181-193.

25. Zhao, W.; Ghorannevis, Z.; Amara, K. K.; Pang, J. R.; Toh, M.; Zhang, X.; Kloc, C.; Tan, P. H.; Eda, G. Lattice dynamics in mono- and few-layer sheets of $WS_2$ and $WSe_2$. *Nanoscale* **2013**, *5*, 9677-9683.

26. Zhang, X.; Qiao, X.-F.; Shi, W.; Wu, J.-B.; Jiang, D.-S.; Tan, P.-H. Phonon and raman scattering of two-dimensional transition metal dichalcogenides from monolayer, multilayer to bulk material. *Chem. Soc. Rev.* **2015**, *44*, 2757-2785.

27. Xiao, Y.; Zhou, M.; Liu, J.; Xu, J.; Fu, L., Phase engineering of two-dimensional transition metal dichalcogenides. *Sci. China. Mater.* **2019,** *62*, 759-775.

28. Ren, J.; Teng, C.; Cai, Z.; Pan, H.; Liu, J.; Zhao, Y.; Liu, B., Controlled one step thinning and doping of two-dimensional transition metal dichalcogenides. *Sci. China. Mater.* **2019**. Doi: 10.1007/s40843-019-9461-8

29. Lu, X.; Luo, X.; Zhang, J.; Quek, S. Y.; Xiong, Q. Lattice vibrations and raman scattering in two-dimensional layered materials beyond graphene. *Nano Res.* **2016**, *9*, 3559-3597.

30. Zhao, Y.; Luo, X.; Li, H.; Zhang, J.; Araujo, P. T.; Gan, C. K.; Wu, J.; Zhang, H.; Quek, S. Y.; Dresselhaus, M. S.; Xiong, Q. Interlayer breathing and shear modes in few-trilayer $MoS_2$ and $WSe_2$. *Nano. Lett.* **2013**, *13*, 1007-1015.

31. Loudon, R. The raman effect in crystals. *Adv. Phys.* **1964**, *13*, 423-482.

32. Baum, A.; Milosavljević, A.; Lazarević, N.; Radonjić, M. M.; Nikolić, B.; Mitschek, M.; Maranloo, Z. I.; Šćepanović, M.; Grujić-Brojčin, M.; Stojilović, N.; Opel, M.; Wang, A.; Petrovic, C.; Popović, Z. V.; Hackl, R. Phonon anomalies in fes. *Phys. Rev. B* **2018**, *97*, 054306.

33. Granado, E.; García, A.; Sanjurjo, J. A.; Rettori, C.; Torriani, I. Magnetic ordering effects in the raman spectra of (formula presented). *Phys. Rev. B* **1999**, *60*, 11879-11882.

34. Jin, W.; Kim, H. H.; Ye, Z.; Li, S.; Rezaie, P.; Diaz, F.; Siddiq, S.; Wauer, E.; Yang, B.; Li, C.; Tian, S.; Sun, K.; Lei, H.; Tsen, A. W.; Zhao, L.; He, R. Raman fingerprint of two terahertz spin wave branches in a two-dimensional honeycomb ising ferromagnet. *Nat. Commun.* **2018**, *9*, 5122.

35. Yao, X.; Ma, J.; Lin, Y.; Nan, C.-w.; Zhang, J., Magnetoelectric coupling across

the interface of multiferroic nanocomposites. *Sci. China. Mater.* **2015,** *58*, 143-155.

AUTHOR INFORMATION

**Corresponding Author**

*Email: bo_peng@uestc.edu.cn ;

*Email: zhanglinbo@uestc.edu.cn

**ORCID**

Bo Peng: 0000-0001-9411-716X

**Author Contributions**

B.P developed the concept, designed the experiment and prepared the manuscript. Y.C.C. synthesized the $CrI_3$ crystal. B.W.D, K.G., Z.L, C.F.G and Z.T.S prepared the $CrI_3$ samples and performed the Raman measurements. L.B, P.H.Z, L.Z, H.P.L and L.B.Z contributed to mechanism of Raman scattering.

**Acknowledgment**

We acknowledges financial support from National Science Foundation of China (51602040, 51872039), Science and Technology Program of Sichuan (M112018JY0025) and Scientific Research Foundation for New Teachers of UESTC (A03013023601007).

**Competing interests:** The authors declare no competing financial interests.

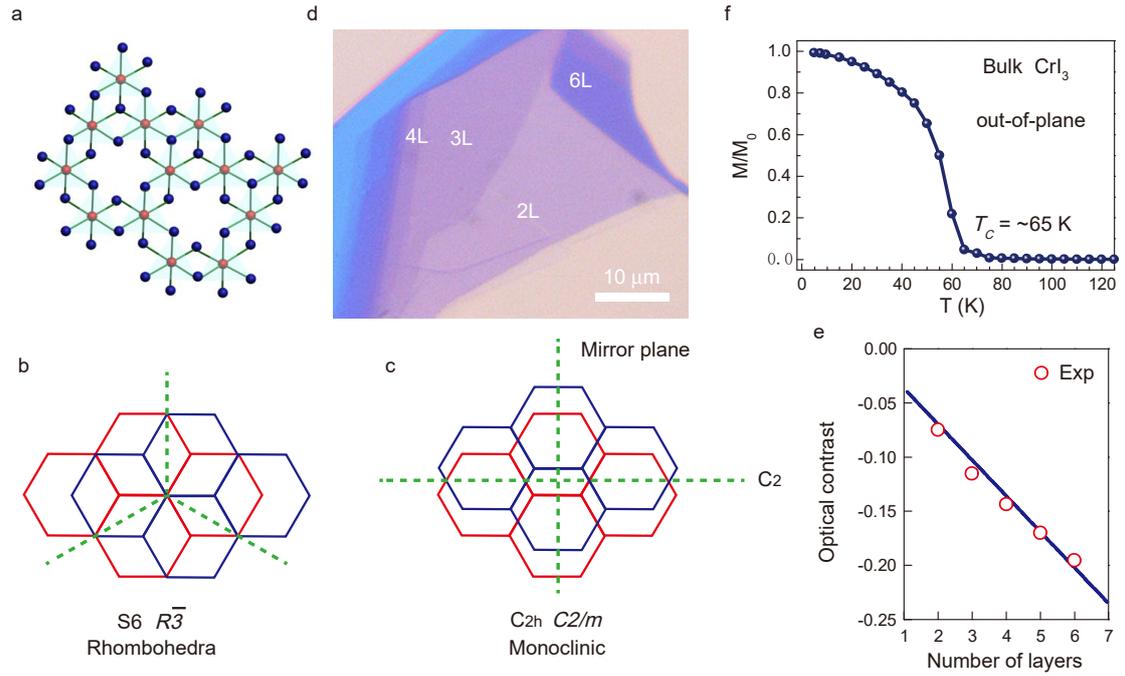

**Figure 1** Exfoliated few-layer CrI$_3$. (a) Atomic structures of monolayer CrI$_3$. (b, c) Rhombohedral (b) and monoclinic (c) stacking order in bilayer CrI$_3$. The rhombohedral structure has an out-of-plane C$_3$ axis and a symmetric center ($S_6 = C_3 + i$), while the monoclinic structure has an in-plane C$_2$ axis and a mirror plane. (d) Optical micrograph of the exfoliated few-layer CrI$_3$. (e) Optical contrast of the CrI$_3$ samples with different numbers of layers (red circles). The blue solid line is calculated results based on Fresnel's equations from Xu *et al.* (Ref. 1). (f) Magnetization of bulk CrI$_3$ as a function of temperature, indicating that $Tc$ = ~65 K.

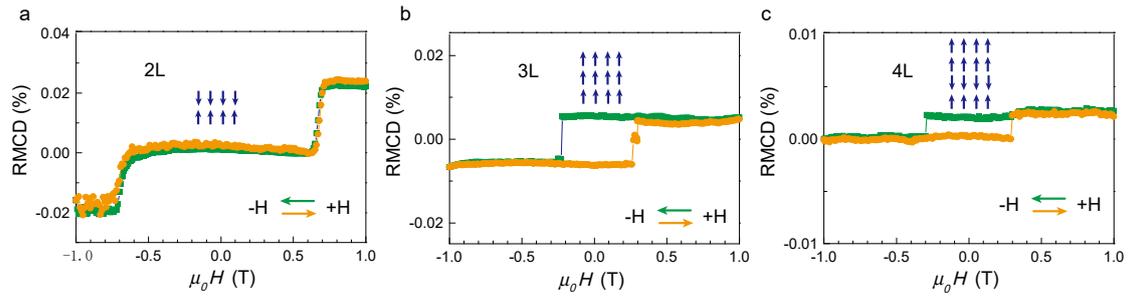

**Figure 2** Layer-dependent magnetic ordering in atomically-thin $CrI_3$ at 10 K. RMCD signal on a 2L (a), 3L (b) and 4L (c) $CrI_3$ flake, showing antiferromagnetic behavior in bilayer $CrI_3$ and ferromagnetic behavior in 3L and 4L $CrI_3$. The blue arrows indicate the magnetization the magnetization orientation in different layer.

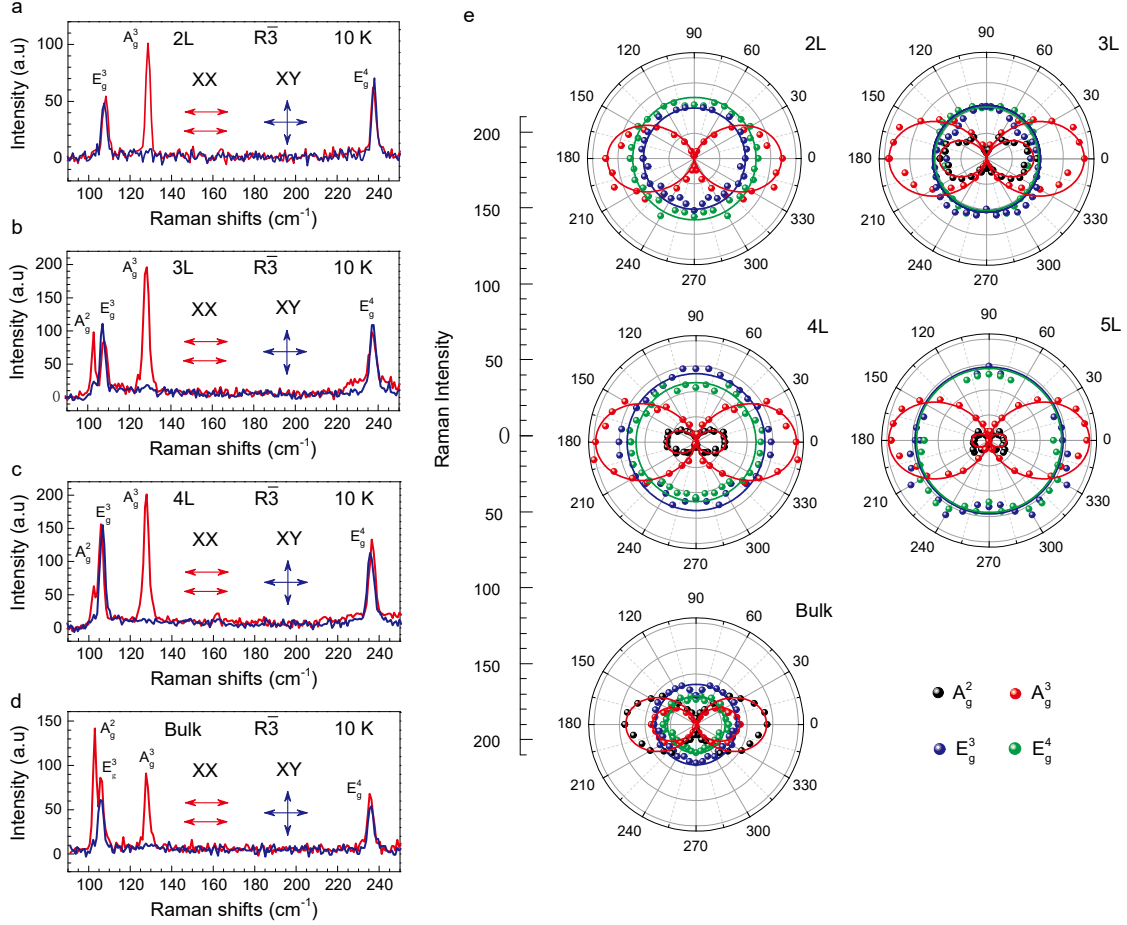

**Figure 3** Layer dependence of the stacking orders in CrI$_3$ at 10 K. (a-d) Raman spectra of 2L (a), 3L (b), 4L (c) and bulk CrI$_3$ (d) in the parallel (XX, red curve) and perpendicular (XY, blue curve) polarization selection channels. (e) Polarization angle dependence of Raman intensity in 2-5 layer and bulk CrI$_3$. The Raman feature at ~107 and ~234 cm$^{-1}$ is degenerated $E_g^3$ and $E_g^4$ mode for the Rhombohedral phase, whose intensity of are independent of the polarization angle; in contrast, the $A_g^2$ and $A_g^3$ modes at ~103 and 128 cm$^{-1}$ show distinct a two-fold polarization dependence.

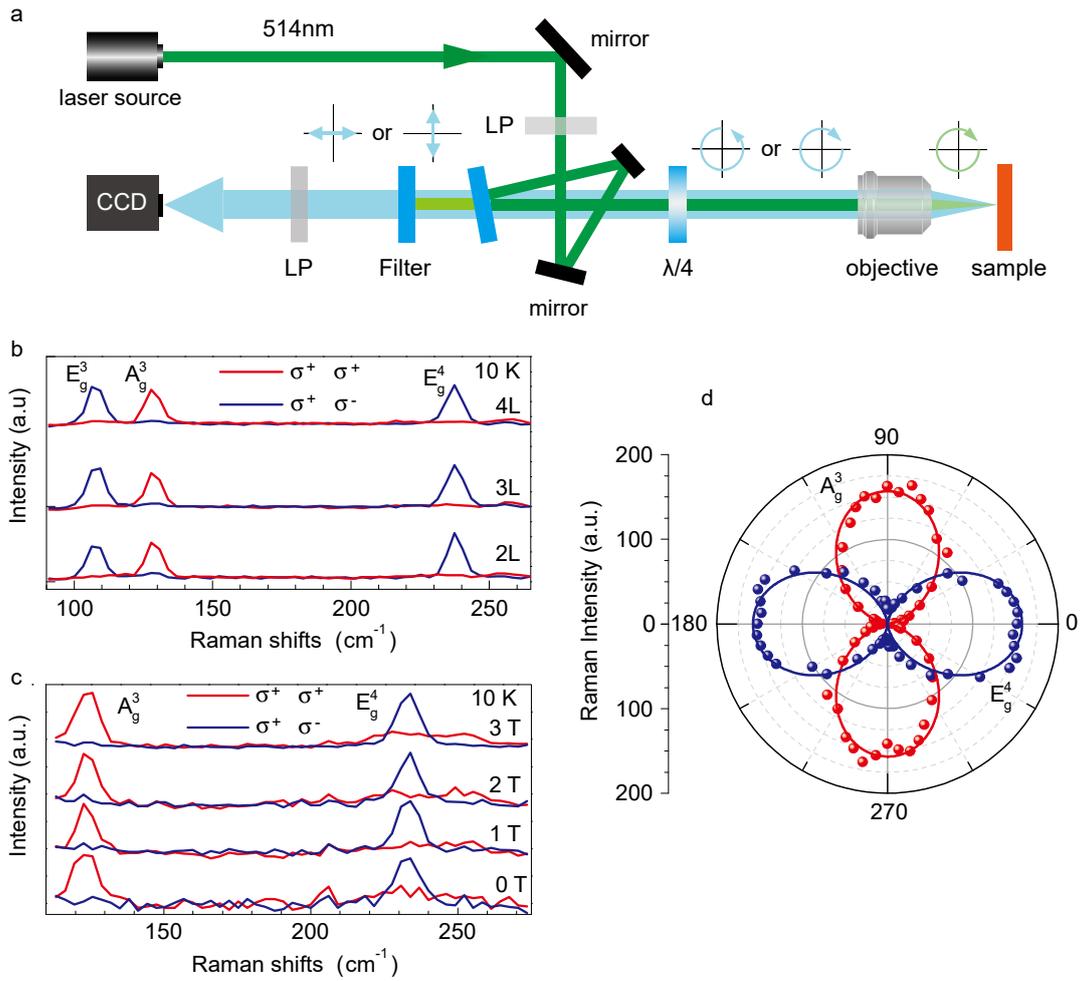

**Figure 4** Helicity-resolved Raman spectra at 10 K. (a) Schematic of the helicity-resolved experimental optical setup. (b) Helicity-resolved Raman spectra of 2L, 3L and 4L CrI$_3$. (c) Helicity-resolved Raman spectra of 3L CrI$_3$ as a function of the magnetic field. (d) Corresponding angular dependence of the $A_g^3$ and $E_g^4$ mode intensities for 3L CrI$_3$, showing a distinctly opposite helicity.

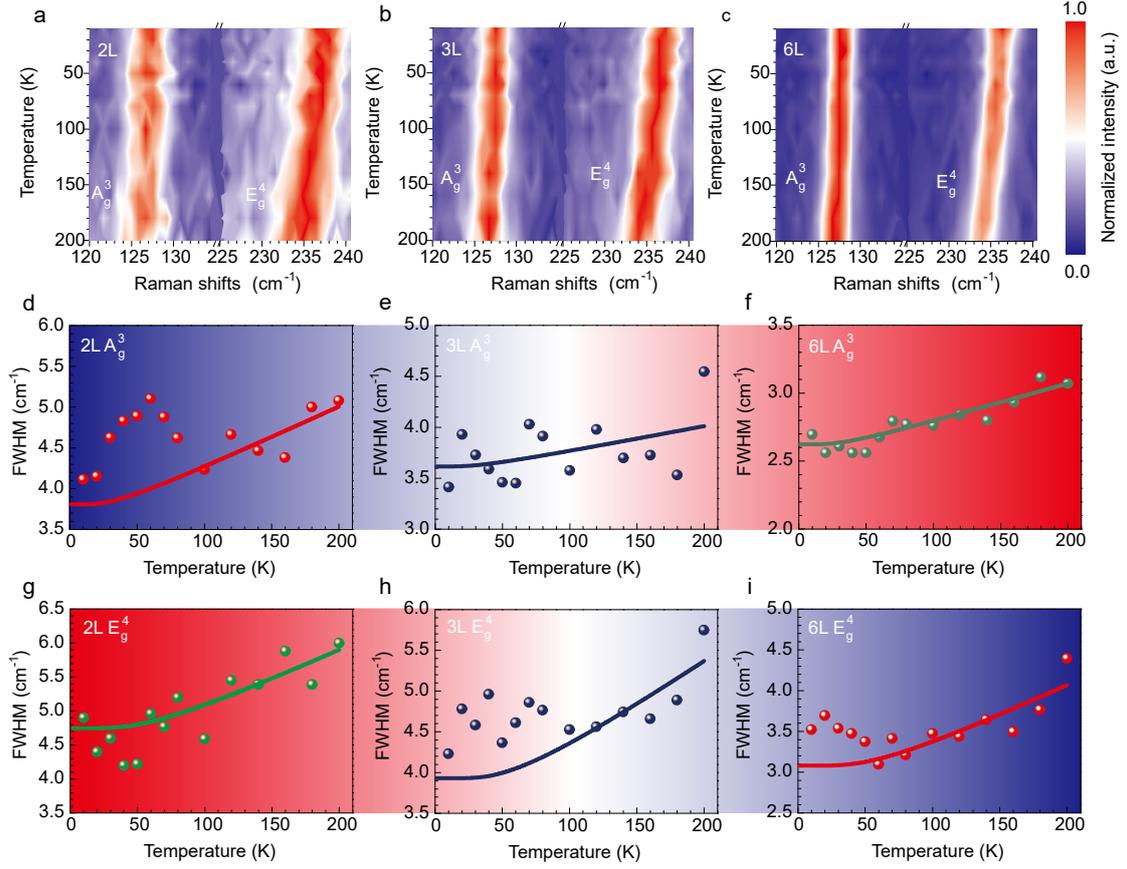

**Figure 5** Temperature dependence of the $A_g^3$ and $E_g^4$ Raman modes. (a-c) Normalized Raman intensity maps of the 2L, 3L and 6L CrI$_3$ as a function of temperature. The Raman features shift to higher frequencies with decreasing temperatures. (d-i) The linewidths of the $A_g^3$ and $E_g^4$ Raman features as a function of temperature as extracted from Fig. 3a-b. The sudden increases in the linewidths indicate the occurrence of spin-phonon coupling.

TOC:

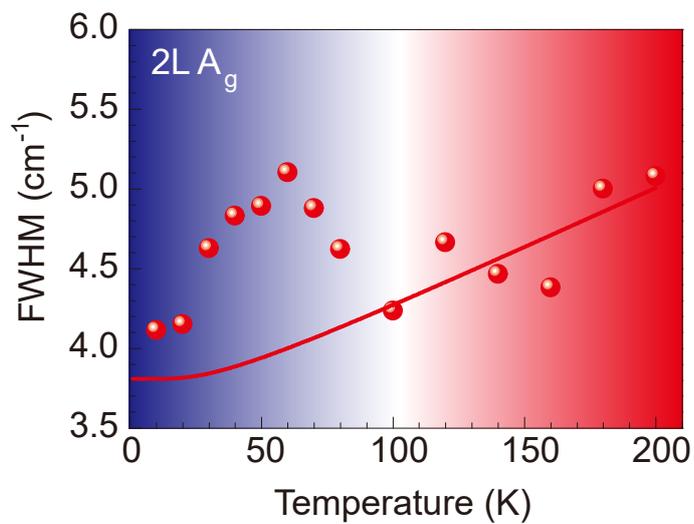

We study the interlayer coupling and stacking order of non-encapsulated 2-5 layer and bulk $CrI_3$ at 10 K; demonstrate a rhombohedral stacking in non-encapsulated $CrI_3$, rather than monoclinic stacking. The spin-phonon coupling takes place below 60 K, resulting in an anomalous temperature dependent deviation behavior.